\documentclass[preprint,12pt,authoryear]{elsarticle}

\usepackage{amssymb}
\usepackage{bbm}
\usepackage{amsmath}
\usepackage{amsfonts}

\journal{Journal of Archaeological Science}

\usepackage[dvipsnames]{xcolor}
\usepackage{datatool}

\usepackage{booktabs} 
\usepackage{array} 
\usepackage{graphicx} 
\usepackage{subfig}

\begin{document}
\frenchspacing

\begin{frontmatter}



\title{Bayesian inference with Monte Carlo approximation: Measuring regional differentiation in ceramic and glass vessel assemblages in Republican Italy,\\ ca. 200 BCE -- 20 CE}


\author{Stephen A. Collins-Elliott}

\address{1101 McClung Tower, Department of Classics, University of Tennessee, Knoxville, TN 37996-0413, sce@utk.edu}

\begin{abstract}
Methods of measuring differentiation in archaeological assemblages have long been based on attribute-level analyses of assemblages. This paper considers a method of comparing assemblages as probability distributions via the Hellinger distance, as calculated through a Dirichlet-categorical model of inference using Monte Carlo methods of approximation. This method has application within practice-theory traditions of archaeology, an approach which seeks to measure and associate different factors that comprise the \textit{habitus} of society. It is implemented here focusing on the question of regional food consumption habits in Republican Italy in the last two centuries BCE, toward informing a perspective on mass social change.

\end{abstract}


\end{frontmatter}

\section{Introduction}

The aim of this paper is to provide a sound, quantitative method for estimating the degree of differentiation in the habits of eating and drinking thorugh mass datasets of vessel assemblages, in order to examine the question of cultural change in Italy leading up to and under the Augustan revolution. While the reign of Augustus has long been seen as a major watershed in cultural change in Italy, especially in the proliferation of a homogenous and unified iconography and style centered on the imperial family \citep{zanker_power_1988,galinsky_augustan_1996}, pursuant to the rise of Roman imperialism \citep{crawford_italy_1996,torelli_tota_1999,keay_italy_2001,bradley_romanization:_2007}, investigation into changes in everyday norms has not received as much  attention.  This paper thus implements a model to measure and detect cultural change in one specific behavior, that of eating and drinking, using a sample set of archaeological vessel assemblages from three different regions in Italy---Etruria, Latium, and Apulia---in order to analyze changes in mass society through habits of food consumption leading up to and under the Augustan regime, from ca. 200 BCE to 20 CE. 

Previous approaches to the comparisons of vessels have largely relied on attribute-level analyses of similarity and difference \citetext{\citealt[136-139]{doran_mathematics_1975}; \citealt[96]{morgan_pots_1991}}, or \textit{ad hoc} impressions of percentages of different classes of ceramics. More recently, correspondence analysis (CA) has proven both popular and effective at comparing archaeological assemblages \citetext{\citealt{cool_peeling_1999,lockyear_site_2000,lockyear_applying_2013}; \citealt[136-146]{baxter_statistics_2003}; \citealt{baxter_correspondence_2010}}. Indeed, the appearance of CA was concomitant with the development of multivariate approaches in the field of sociology, to identify regimes of taste and sort out the different \textit{habitus} which comprise society \citep{benzecri_analyse_1973,bourdieu_anatomie_1976,bourdieu_distinction_1979}. Today, CA has  been largely separated from its post-structuralist origins  \citetext{but see \citealt{pitts_artefact_2010}}. Yet, if archaeological assemblages are nothing other than the material waste of a systemic context, once living, and if indeed material factors (like taste in art) were an inherent part of research into the differentiation of \textit{habitus} among late 20th-century sociologists using multivariate techniques \citep{lebaron_how_2009}, then the use of multivariate analyses like CA in establishing patterns of lifestyles is not only a sensible, but by now conventional, undertaking in archaeology.

What is treated here is the exploration of a new model, focusing not on the total construction of \textit{habitus} writ large but rather aimed at measuring the degree of difference in one particular set of behaviors, eating and drinking. It views the sum of vessels as a set of proxy evidence for these behaviors, but includes not just those vessels which were used in the immediate act of consumption, like tableware, but vessels which speak to the contents of those vessels---wares and forms which pertain to modalities of storage, transport, and preparation. These were treated further as background information which played a part in constructing the manners of food and drink consumed in their society, included in order to yield as information-rich a picture of the instance of consumption as possible.

In other words, the ceramic and glass vessels which survive as after-effects of the instance of consumption serve to provide a ``snapshot'' of a meal which took place in the past. These artifacts are moreover not just reflective of norms and practices, but constitutive of them as well  \citep[701-704]{pitts_emperors_2007}. While the  sum of  these meals  are unknowable as an experience,  the different categories of vessels which were used  in them nevertheless allow for the reconstruction of differences in the overall habits of their users, as well as the economic and social systems which implicated them, in order to look at broader social change.

The development of a new model for measuring such changes in habit is merited for several reasons. First are the factors of time and rate of change: when examining the change in habits diachronically, comparisons of assemblages using a technique like CA would necessitate an entire series of biplots, one for each time interval, whose measurement from interval to interval could prove vexing. Furthermore, the chi-squared distance, on which correspondence analysis is based, necessitates (strictly speaking) counts of data, rather than frequencies \citep[258]{cowgill_selection_1975}, and the fact that the chi-squared distance is not symmetric means that one assemblage cannot be directly compared with another. Finally, it is desirable to have a credible region which indicates the level of certainty about the conclusions that reflects the sample size of the data used.

Accordingly, this paper illustrates the implementation of a probabilistic approach that can address these issues. The model presented here is one founded upon a Bayesian view of probability as uncertainty, which expresses credibility in its results directly through the observation of  data \citep{buck_bayesian_1996,robertson_spatial_1999,lindley_philosophy_2000}. In informal terms, the method offered in this paper is based on the ``urn-draw'' model of probability problems, in which each instance of a type of archaeological find is treated like a draw of a different colored object from an urn. An assemblage is just the collection of draws from an urn, or, taking into account the variable of time, a process which is conceived of draws spread out over an entire set of urns, one representing each time interval. Rather than comparing these sets of draws from their respective urns in total after their collection, this model makes a comparison between sets after each draw, which allows us to incorporate information about the effects of sample size into the estimation of the degree of difference. Because the order of draws affects these estimations, it is permuted to provide more information about the most likely range of values for the degree of difference.

In more technical terms, the calculation of difference among two assemblages is obtained through conceiving of each time interval's artifact assemblage as a probability distribution, and then the distance between those distributions is calculated using the Hellinger distance. Probability distributions are modeled as the result of a Dirichlet-categorical hierarchical model of inference, with different vessel taxa representing different categories, and with quantities distributed over their potential time intervals using a uniform distribution. In order to obtain a credible region as well as a point estimate, the Hellinger distance was calculated after each inferential step of the Dirichlet-categorical hierarchy. Owing to the non-exchangeability of the likelihood for computing the Helligner distance, Monte Carlo methods were used to permute the order of each inferential step toward approximating the Hellinger distance. From these Monte Carlo values, it was possible to find a point estimate for the most probable value, represented by the mode, and the credible region, defined by the highest posterior density  around the mode.

\section{Data Collection}

\begin{figure}[t!]
		\centering
		\includegraphics[width=1\textwidth]{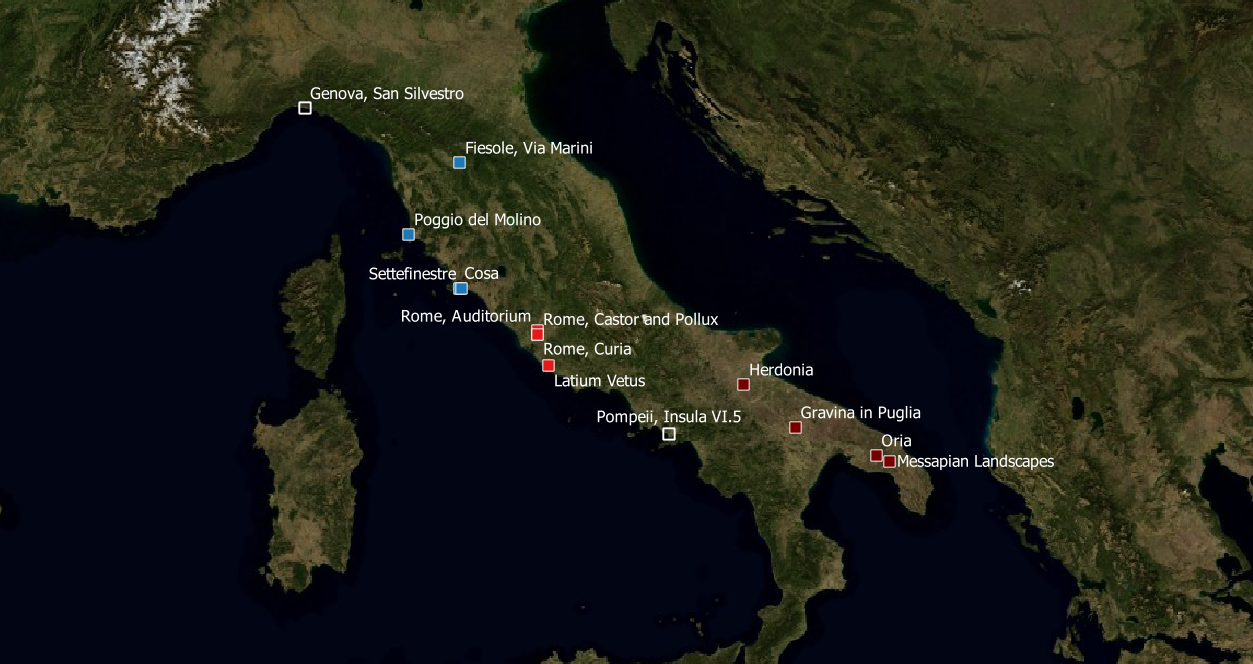}
		\caption{Map of the fourteen sites sampled for the algorithm, color-coded by their region (Etruria, Latium, and Apulia).\label{jasmap}}
	\end{figure}

In order to illustrate the theoretical approach of this study, I selected the published results of 14 archaeological projects that contained vessel finds datable to the last two centuries BCE  (Table \ref{bibl1}). Information on the database  used in this article can be found in \ref{appendix1}.\begin{table}[t!]
	\centering
	\scalebox{0.85}{
		\begin{tabular}{cll}
			\toprule
			{Key} & {Name} & Reference \\
			\midrule
			{101}	&	Genova	&\cite{milanese_genova_1993} \\
			{201}	&	Settefinestre	& \cite{ricci_settefinestre._1985}\\
			{210} 	&	Cosa &	\cite{fentress_cosa_2003}\\
			{220}	&	Fiesole, Via Marini &	\cite{de_marinis_archeologia_1990} \\
			{221}	&	Poggio del Molino&	\cite{de_tommaso_villa_1998}\\
			{301}	&	Rome, Auditorium	&\cite{carandini_fattoria_2006}\\
			{309}	&	Rome, Curia &	\cite{morselli_curia_1989}\\
			{315}	&	Rome, Temple of Castor and Pollux	&	\cite{bilde_temple_2008}\\			
			{350}	&	Latium Vetus (Nettuno) Survey &	\cite{attema_between_2011}\\
			{401}	&	Pompeii, Insula VI.5 	&	\cite{bonghi_jovino_ricerche_1985}\\
			{505}	&	Herdonia	& \cite{de_stefano_contesto_2008}\\
			{610}	&	Oria	&	\cite{yntema_search_1993}\\
			{612}	&	Messapian Landscapes &	\cite{burgers_constructing_1998}\\
			{680}	&	Gravina in Puglia		& \cite{small_gravina._1992-1} \\
			\bottomrule
		\end{tabular}}
		\caption{Bibliographic citations of publications whose vessel data served to provide the pilot dataset.\label{bibl1}}
	\end{table} The sites, located in pensinsular Italy (Fig. \ref{jasmap}), were chosen on the basis of the quality of the data published,  such that they included information on the classification, typology, quantity, and dating of different classes of vessels in a balanced manner. Publications which consist of a study of a single class of material were not included, owing to their potential tendency to over-represent that class in the absence of other classes of materials.
	
These projects  provide a sample of evidence for the sake of illustrating the model. The tendencies of the publications to represent largely urban centers or rural estates are the disciplinary tendencies of classical archaeology.  Combining surface survey and excavated material was likewise a deliberate choice, to experiment with the sliding scale of defining context in archaeology. The lens was thus set as broadly as possible, to encompass a regional level, in order to illustrate the method of normalizing counts of data (see Section \ref{sec:context}). Because context can be scaled and adjusted at hand to suit the needs of a particular investigation at a broad regional level, imposing the condition that only datable finds from primary deposits should be included in the study would unnecessarily ignore a significant amount of information. 

Sampling has been a persistent topic of discussion in quantitative approaches to archaeology, and has been treated at depth with especial regard to the subject of context, above all in surface survey  \citetext{\citealt{cowgill_selection_1975}; \citealt{nance_regional_1983};  \citealt[361--400]{shennan_quantifying_1997};  \citealt{orton_plus_1999,orton_sampling_2000};  \citealt[38--49]{baxter_statistics_2003}; \citealt{terrenato_sample_2004}}.  Yet, inherent in the sampling of archaeological finds is the fact that the actual sample of material in its depositional context lies beyond the control of the investigator. This is where a Bayesian approach to data analysis becomes useful, as it does not place the same emphasis on sampling criteria that a frequentist ``test'' would. That said, it would be incorrect to assert that the method of data collection has no bearing on results obtained through Bayesian inference \citep[197-199]{gelman_bayesian_2014}. While the data collection for this study consisted of published reports, it is nevertheless a problem in archaeology that the sample size of finds to be collected is beyond the control of the archaeologist: the finds in a deposit or topographic unit are dictated by an unknowable sum of systemic and formation processes, and it is not possible to collect more samples if one has collected all the finds in a context. Thus, it is necessary to bring to bear sample size on the results of one's inferences, in order to establish levels of certainty in the results. 

In categorizing the ceramic and glass vessel finds, two different parameterizations were used. The first consisted of vessel ware or manufacturing technique, to illustrate differentiation in exterior vessel style. The second comprised vessel form, or morphology to emphasize function, which can be taken as abstractions of the way in which the ancient inhabitants of site transported, prepared, and consumed their food. While each of these two parameterizations, form and ware, can be taken as a check on another, they can also be taken to emphasize certain aspects of the habits of eating and drinking: vessel form after all draws greater emphasis to the function or use of vessels, while vessel ware pertains to non-functional elements of style or appearance, additional criteria which adds a significant level of  texture to the description of assemblages. It should be acknowledged that any manner of parameterization is possible on the same set of material, in whatsoever way the critical aims the vessel taxonomy should warrant.

\subsection{Synthetic Categorization}

Owing to the multiple and often overlapping classifications and typologies developed for Republican and Augustan Italy, a standardized set of categories was developed in order to employ a uniform taxonomy. That is, it was not only necessary to digitize vessel quantities from the published results of older projects, but it was also necessary to establish a concordance to render finds comparable with one another. Categorical assignment of the vessels was accordingly divided between traditional descriptions of vessel form (morphological-functional) and ware (technical-functional) using {\texttt{synthkat}, a taxonomic algorithm designed to translate across different methods of organizing vessel assemblages \citep{collins-elliott_agglomerative_2016}}.  The categorization (whether classification or typology) of the vessel was described in semantic sets, in two different parameterizations,  the form and ware of the vessel. The resulting parameterizations can be found in Table \ref{parameters}.
	
	The aim was to collect all those vessel forms and wares which could be used in the production, transportation, storage, and consumption of food.  To be sure, many of these vessels could be used outside of the experience of eating and drinking---lamps, which point to differentiation in the use of artificial light (as an alternative or accompaniment to a hearth) being the most obvious. But the same is true even when considering food itself: foodstuffs can be eaten and also used in artisanal or industrial activities, like olive oil. Determining which are which in the archaeological record would be a frustrating task, and these categories should be treated as a form of approximation, rather than a strictly defined set of variables. The approach taken here has sought to be as comprehensive as possible in the selection of vessels, to create an abstract picture of dining practices using the collective vessel waste of a community. Other methods relying solely on specific classes of vessels are likewise feasible, such as examining burn patterns on cookware \citep{campanella_il_2008,banducci_function_2014}.

	\begin{table}[t!]
		\centering
		\begin{tabular}{ll}
			\toprule
			Vessel Form       &Vessel Ware       \\
			\midrule
			transport amphora & black gloss       \\
			beaker            & lamp              \\
			bowl              & transport amphora \\
			plate             & Pompeian red slip \\
			lid               & thin-walled       \\
			{olla}            & grey              \\
			bottle, pitcher   & terra sigillata   \\
			basin             & grey gloss        \\
			mortarium         & unslipped, fine   \\
			pot               & unslipped, coarse \\
			pan               & cookware          \\
			storage jar       & red gloss         \\
			casserole         & glass             \\
			skyphos           & lead-glazed       \\
			                   kylix        &     \\
			                   krater        &    \\
			                   lamp &	\\
			                  \bottomrule
		\end{tabular}
		\caption{Parameterizations for vessel form  and vessel ware  in the database (\ref{appendix1}).\label{parameters}}
	\end{table}

Regarding the method of vessel quantification, sherd count was used out of practical considerations (see comments in \ref{appendix1}). While other and indeed multiple methods of quantification would be desireable, like estimated vessel equivalents, it should be acknoweldged that the use of sherd count alone is a biased estimate of vessel quantification, since it is only unbiased in the case when comparing across two different vessel classes which will have the same breakage rates \citep[31]{orton_quantitative_1975}. That said,  the use of sherd count alone for this particular study can be validated in so far as the goal here is not quantification in itself, nor the estimation of a proportion of a particular class in comparison to another class, but rather the comparison of entire assemblages to one another. In the case of parameterization by vessel ware, the same bias will hold in the case of each particular class, thus resulting in comparisons of two assemblages which are subject to the same biases. In the case of parameterization by vessel form, the same assumption is weakened by the fact that vessel morphology does not necessarily correspond to its manufacture (and thus its breakability).

Some additional comments can however nuance this weakness. Transport amphorae, one of the most influential categories, have a significantly more robust fabric than tableware and thus lower breakability rate, and thus remain in their own functional category. The main detriment lies in the fact that glass has a higher breakability rate than ceramic, and thus the influence of its fragmentation underlies many other functional categories under the parameterization by vessel form, although according to the parameterization by vessel ware, glass shards had a minimal presence in the database. To give an idea, their shard count amounted to a total of 329 (in 129 records) out of a total count of 33,047 (by vessel form and ware) in the database, and thus amounts to a small fraction of any possible influence. Further work can however be done to shed light on the breakbility rates of the different functional categories of tableware.

	\section{Modeling the problem: Hellinger distances between Dirichlet-categorical distributions}


The process of quantifying archaeological finds can be likened to an urn problem, a standard model for probabilistic problems which have the goal of estimating the frequency of different colored object in an urn (Table \ref{urnexample1}). An individual selects one object at a time in order to gain an idea about the total composition within the urn, which is unknown. While some urn problems involve sampling with replacement (returning the object to the inside of the urn prior to the next draw), the recovery of archaeological finds is more akin to sampling without replacement.  Thus, the model proposed here can be applied to any quantity of ceramic vessel fragments that have been excavated from the ground (as reported in the published datasets), belonging to a particular context (however broadly or narrowly defined).

There are several preliminary considerations which should be taken into account regarding the model. First is the need for computability. For example, the hypergeometric distribution, which is typicaly used to model the process of sampling without replacement \citep[79-84]{johnson_urn_1977}, is impracticable with large numbers of finds in real-world applications. Second, there should be a way to incorporate the sampling size into the degree of certainty about the estimates obtained from the urn draws. Third, it is important to bear in mind that the value of interest is not the frequencies of different categories of vessels (the frequency of different colored objects in the urn), but rather the \textit{degree of difference} between two such assemblages (between two urns).

In other words, the aim is to find a distance $\phi$ between any two probability distributions $p$ and $q$. On the one hand, a typical approach would be to collect data and then to measure the distance between the two distributions. Yet, any distance measure would not reflect the level of certainty inherent in the sampling size of either distribution. This is where calculating $\phi$ through a framework of Bayesian inference can help. With categorical data, like the different categories of ceramics (in the urn model, the different colors of objects), we can create a model of hierarchical inference as follows. For a set of parameters $\theta_i$, where $i = 1, \ldots, K$, for $K$ different categories of objects, there is a corresponding set of observable data, $x_i$. Let $\boldsymbol{\theta}$ and $\boldsymbol{x}$ be the respective vectors of the parameters and data.  Bayes' theorem states that a posterior probability, $p(\boldsymbol{\theta|x})$, is proportional to a prior probability, $p(\boldsymbol{\theta})$, times a likelihood, which is represented as $p(\boldsymbol{x|\theta})$, the value of the parameter $\boldsymbol{\theta}$ based on the data $\boldsymbol{x}$ \citep[144-5]{buck_bayesian_1996}. In formal terms, $p(\boldsymbol{\theta|x}) \propto p(\boldsymbol{x|\theta}) p(\boldsymbol{\theta})$.

The observation of data consists of a categorical distribution, representing the event of the draw from the urn, $p(\boldsymbol{x|\theta}) \sim \mathrm{Cat}(K, \boldsymbol{\theta})$. The categorical distribution can be denoted $p(x=i|{\boldsymbol {\theta}})=\theta_{i}$, which describes an event where there are $K$ mutually exclusive results, such that if category $i$ is the outcome,  $x_i = c_i$, where $\boldsymbol{c} = (c_1, \ldots, c_K)$, is the set of the observed quantity from the likehood, $c_i = \{i : x_i = i\}$, and $\sum_{i=1}^K \theta_i  =1$. In the case of an urn draw, this typically assumes that $x_i = 1$ and all other values are $0$.  The observation of that draw will change our prior estimate, $p(\boldsymbol{\theta})$, leading us to a new posterior estimate $p(\boldsymbol{\theta|x})$. The Dirichlet distribution, defined as
\begin{equation*}
 \mathrm{Dir}(\boldsymbol{\alpha}) \sim \frac{\Gamma\left(\sum_{i=1}^K\alpha_i\right)}{\prod_{i=1}^K{\Gamma}(\alpha_i)} \prod_{i = 1}^{K} \theta_i^{\alpha_i - 1},
\end{equation*}

 where $\boldsymbol{\alpha} = (\alpha_i, \ldots, \alpha_K)$ is used as a conjugate prior distribution to compute the posterior  analytically \citetext{\citealt[578--9]{gelman_bayesian_2014}; \citealt[133, 328--31]{buck_bayesian_1996}}. Following Bayes' theorem, the posterior probability distribution will be
\begin{align*}
p(\boldsymbol{\theta|x}) & \propto  \mathrm{Cat}(K, \boldsymbol{\theta}) \mathrm{Dir}(\boldsymbol{\alpha})\\
& = \prod_{i=1}^{K} \theta^{\left(x = i\right)} \prod_{i = 1}^K \theta^{\alpha_i - 1}\\
& = \prod_{i=1}^{K} \theta^{c_i} \prod_{i=1}^{K} \theta^{\alpha_i -1} = \prod_{i=1}^{K} \theta^{c_i + \alpha_i - 1}\\
\end{align*}

Thus, $p(\boldsymbol{\theta|x}) \sim \mathrm{Dir} (\boldsymbol{\alpha + c})$.  the posterior distribution itself being a Dirichlet distribution, the expected value of a parameter in the posterior will have the form
\begin{equation*}
E[\theta_i]  = \frac{\alpha_i + c_i}{\sum_i \alpha_i + n}.
\end{equation*}
where $n$ = $\sum_i c_i$, for each category $i = 1, \ldots, K$. 

Notwithstanding the cumbersome definitions, the intuition behind the model is straightforward: estimating the degree of difference between assemblages after each draw from the urn.
Thus, let two urns contain\ $K$ different colored objects, which are drawn from the urn $N$ number of times.

The first draw  results in a posterior probability distribution. This posterior distribution can be treated as a new prior probability, and the Dirichlet-categorical inference repeated in a hierarchical chain for as long as one is conducting draws from the urn. Table \ref{urnexample1} illustrates the hierarchical Dirichlet-categorical model at work for an example of three urns, each of which contains three different colors of objects ($K = 3$, with red, green, and blue colored objects), where there are ten draws made from each urn ($N = 10$). One can note that after ten draws, urns 1 and 2 are identical in terms of the frequency of their categories (they each contain 1/2 red, 4/5 green, and 1/10 blue items), even though the order of the selection of the objects differed. The collected frequencies of different colors represented in each urn is a probability distribution, an estimate of their actual frequency inside the urn.\begin{table}[t!]
	\centering
	\scalebox{0.85}{
	\begin{tabular}{*{15}{c}}
		\toprule
		& \multicolumn{4}{c}{Urn 1} &	& \multicolumn{4}{c}{Urn 2} &	& \multicolumn{4}{c}{Urn 3}  \\
		\cmidrule(lr){2-5} \cmidrule(lr){6-10} \cmidrule(lr){11-15}
		Draw	&	R	&	G	&	B	&		& &	R	&	G	&	B	&		& &	R	&	G	&	B	&		\\
		\midrule
		1	&	1.00	&	0.00	&	0.00	&	R	& &	0.00	&	0.00	&	1.00	&	B	& &	0.00	&	0.00	&	1.00	&	B	\\
		2	&	0.50	&	0.50	&	0.00	&	G	& &	0.00	&	0.50	&	0.50	&	G	& &	0.00	&	0.00	&	1.00	&	B	\\
		3	&	0.67	&	0.33	&	0.00	&	R	& &	0.33	&	0.33	&	0.33	&	R	& &	0.33	&	0.00	&	0.67	&	R	\\
		4	&	0.75	&	0.25	&	0.00	&	R	& &	0.50	&	0.25	&	0.25	&	R	& &	0.00	&	0.25	&	0.50	&	G	\\
		5	&	0.60	&	0.20	&	0.20	&	B	& &	0.60	&	0.20	&	0.20	&	R	& &	0.00	&	0.20	&	0.60	&	B	\\
		6	&	0.67	&	0.17	&	0.17	&	R	& &	0.67	&	0.17	&	0.17	&	R	& &	0.33	&	0.17	&	0.50	&	R	\\
		7	&	0.57	&	0.29	&	0.14	&	G	& &	0.71	&	0.14	&	0.14	&	R	& &	0.43	&	0.14	&	0.43	&	R	\\
		8	&	0.50	&	0.38	&	0.13	&	G	& &	0.63	&	0.25	&	0.13	&	G	& &	0.38	&	0.13	&	0.50	&	B	\\
		9	&	0.44	&	0.44	&	0.11	&	G	& &	0.56	&	0.33	&	0.11	&	G	& &	0.33	&	0.22	&	0.44	&	G	\\
		10	&	0.50	&	0.40	&	0.10	&	R	& &	0.50	&	0.40	&	0.10	&	G	& &	0.40	&	0.20	&	0.40	&	R	\\
		\bottomrule& & 
	\end{tabular}	
}\caption{The Dirichlet-categorical model of inference at work for ten draws from three urns, each with red, green, and blue colored objects. After ten draws each, urns 1 and 2 have the same frequency of objects, but note that the order in which those objects were drawn differs. \label{urnexample1}}
\end{table}

In a hierarchical Dirichlet-categorical model of inference, the order in which the objects were drawn does not affect the posterior estimate, as the data satisfy the criterion of exchangeability. Yet, the frequency of a specific category within a distribution is not the value of interest. The \textit{degree of difference} between the distributions is, which we can denote as $\phi$, and it is equally important to obtain information about the strength of certainty in the estimate of $\phi$. 

\begin{figure}[p!]	
	\centering
			\includegraphics[width=0.6\textwidth]{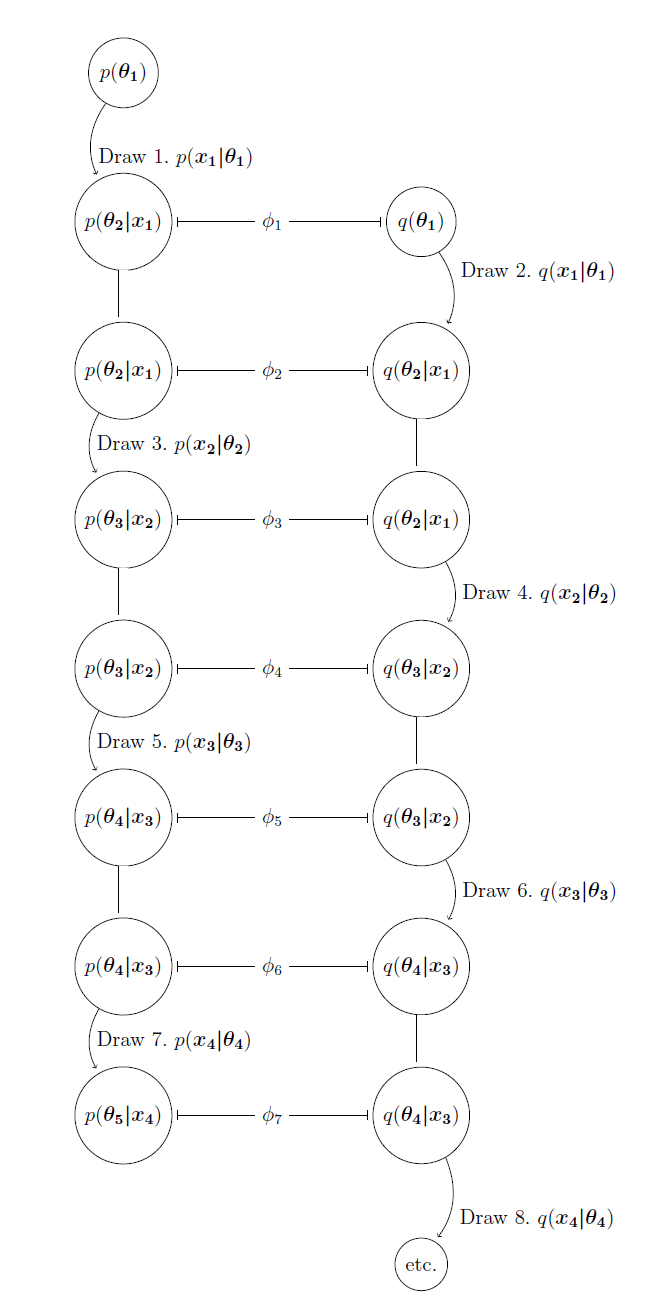}
	\caption{An example of the hierarchical model of Bayesian inference in the case of just two ``urns.'' With each new draw (the likelihood) for the distributions $p$ and $q$, an estimation of the distance metric $\phi \sim H(p,q)$ between the two distributions can be calculated: $\phi_1$ after the first draw, $\phi_2$ after the second, and so on.\label{phimodel}}
\end{figure}

Accordingly, I propose a model where the distance $\phi$ (between any two different probability distributions $p$ and $q$) is calculated after every inference in the hierarchy of the Dirichlet-categorical model, in other words, after every draw from the urn.  To continue the example of the urn model, the result will be a set of values $\phi_1, \ldots, \phi_N$, for $N$ draws from either urn. For measuring this degree of difference between two  probability distributions, $p$ and $q$, I used the Hellinger distance,
\begin{equation*}
H^2(p,q) = \sum_{i=1}^K(\sqrt{p_i}-\sqrt{q_i})^2 = 2 (1-BC(p_i,q_i))
\end{equation*}
where the Bhattacharrya coefficient, $BC(p_i,q_i)$, is defined as $\sum_{i=1}^{K}\sqrt{p_iq_i}$ \citetext{\citealt[442]{gibbs_choosing_2002}; \citealt[61]{pollard_users_2002}}. The Hell\-in\-ger distance results as a scalar value in the set $[0,\sqrt{2}]$, where $0$ indicates complete similarity and $\sqrt{2}$ indicates dissimilarity between the two distributions.\footnote{A\label{note1} second method was originally explored, the Kullback-Leibler divergence \citep{kullback_information_1951}, but as $D_{KL}(p||q) \geq  H(p,q)$ there was no substantial change in the results.} This process is illustrated in Figure \ref{phimodel}.

Thus, to return to the urn example above, we can obtain a series of estimates for the degree of difference between any two urns after every draw. One alternates likelihoods from any two different probability distributions  from each urn, and calculates values of $\phi \sim H(p,q)$ after each draw. In other words, while prior approaches to finds quantification has been essential to collect a sample from each urn, and then to compare afterwards, this approach calculates the distance between assemblages after every draw from the urn. Thus, information regarding certainty owing to sample size can be directly incorporated quantitatively into the estimate of $\phi$. 

\begin{figure}[p!]
	\centering

			\includegraphics[width=0.7\textwidth]{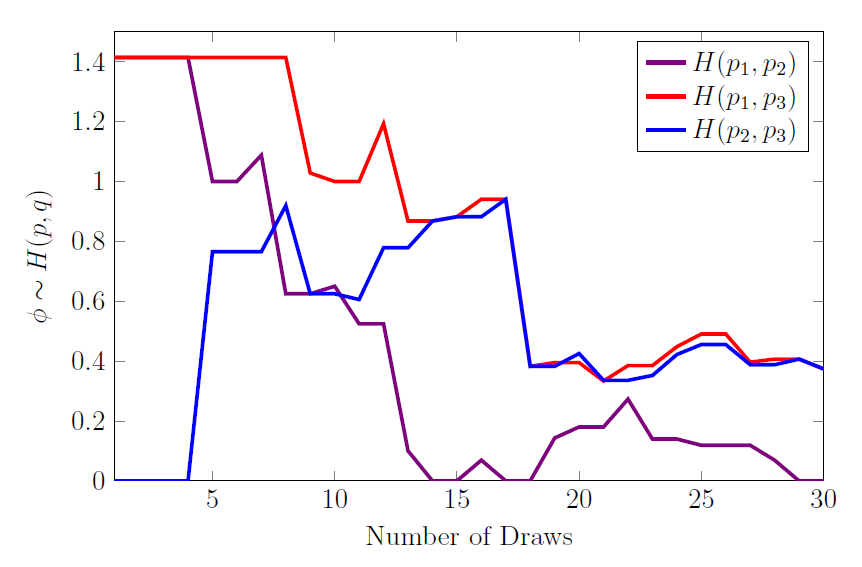}
	\caption{Changing estimates of $\phi$, measuring the degree of difference between each set of urns using the Hellinger distance $H(p,q)$, calculated after every draw between the three urns, from the example in Table \ref{urnexample1}. \label{urnexample3}}	
\end{figure}

\begin{figure}[p!]
	\centering

			\includegraphics[width=0.7\textwidth]{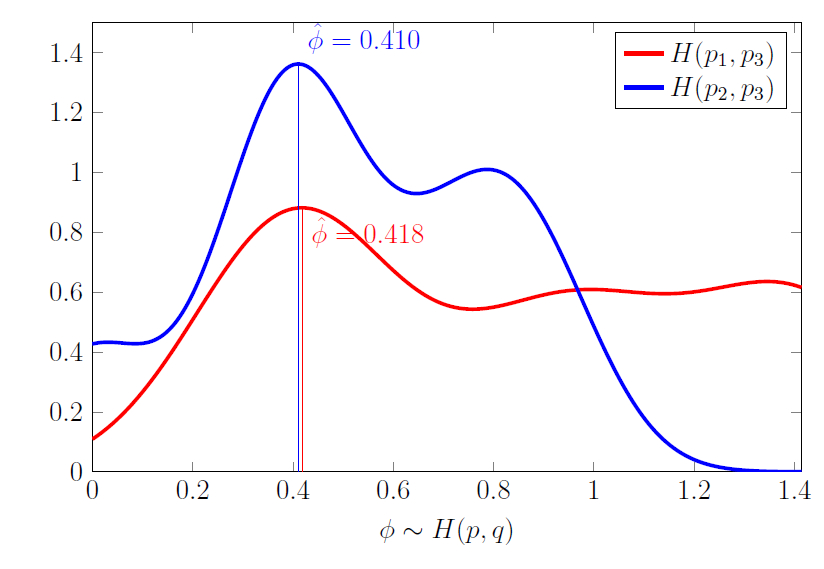}
	\caption{Kernel density estimate of values of $\phi$ for the cases of $H(p_1,p_3)$ and $H(p_2,p_3)$, with the maximum value $\hat{\phi} = \arg\max_{\phi} p(\phi)$, for a set of draws where $N = 30$ (10 draws from 3 urns).\label{kdeexample}}	
\end{figure}

Calculating $\phi$ after every draw serves to provide information about the level of certainty in the difference between the contents of the urns. Given that one could draw a hundred times from one urn, and three times from another, resulting in too small of a sample, it would be desirable to have some knowledge about the effects of the sample size on the estimate of $\phi$. Moreover, one cannot be certain the degree to which the estimate of $\phi$ might be changed if one had stopped collecting, or had gone further in pulling more draws. Returning to the example of the three urns makes this clear.  While $\phi_N(p_1,p_3)$ and $\phi_N(p_2,p_3)$ are equal, at the very end of the process of drawing from the urns--if we had only drawn ten times from all of the three urns, the estimate would be different (Fig. \ref{urnexample3}). Thus, rather than just using $\phi_N$, (that is, rather than collecting data, and then calculating the Hellinger distance), it would be better to use the estimate $\hat{\phi}$, the most frequently returned value of $\phi$ (its mode), which, in the case of $\hat{\phi}(p_1,p_3)$ and $\hat{\phi}(p_2,p_3)$, occurs where $x = 0.418$ for $H(p_1,p_3)$, and at $x = 0.410$ for $H(p_2,p_3)$. This estimate indicates that on the basis of the sampling the urns $p_1$ and $p_2$ are distinct from $p_3$ (Fig. \ref{kdeexample}).

Accordingly, it can also be noted that the order of the draws matters for producing values of $\phi$. On the one hand, the data are exchangeable in a hierarchical model of the Dirichlet-categorical inference. If one was only concerned with a single parameter (a single color represented in the urn, or a single class of vessel), it would not matter which order the data were collected in. On the other hand, changing the order of draws will necessarily change the values of the distance $\phi$. As one can note in Figure \ref{urnexample3}, the frequency of different colors after $N$ draws is identical for both urns, $p_1$ and $p_2$, but the set of the values of $\phi$ with respect to urn $p_3$ are not.

In other words, the set of estimates of $\boldsymbol{\phi} = (\phi_1, \phi_2, \ldots \phi_N)$, are non-exchangeable, which will cause different results for finding the most probable value $\hat{\phi}$: the order of the draws matters for estimating $\hat{\phi}$. In calculating $H(p,q)$ it is certain that in the case of just one permutation of the observation of the data--the likelihood $p(\boldsymbol{x|\theta})$--only the last draw $\phi_N$ will be identical to any previous orders of draws. That is, if the same objects are drawn in different orders, the first which generates one set of estimates,
\begin{equation*}
\phi_1, \phi_2, \phi_3, \ldots, \phi_N
\end{equation*}
and the second which generates other set of estimates, where the order of $p(\boldsymbol{x|\theta})$ in the Dirichlet-categorical model is not identical to the first,
\begin{equation*}
\phi_1^{\prime}, \phi_2^{\prime}, \phi_3^{\prime}, \ldots, \phi_N^{\prime}
\end{equation*}
it will be the case that $\phi_N = \phi_N^{\prime}$, but not  the case that $\phi_1 = \phi_1^{\prime}, \phi_2 = \phi_2^{\prime}$ and so on, due to the change in permutation where at least one $\phi_i \neq \phi_i^{\prime}$. The final estimate of $\phi_N$ should however not be viewed as the most accurate one, since one could just as easily have stopped drawing, or continued to draw.  Accordingly, it should be desirable to use the value $\hat{\phi}$ such that $\hat{\phi} \approx \hat{\phi}^{\prime}$,    as is the case with the sight difference in $\hat{\phi}(p_1,p_3)$ and $\hat{\phi}(p_2,p_3)$ in Figure \ref{urnexample3}.

The problem with the ordering could be circumvented by running every single permutation of the likelihood in the Dirichlet-categorical hierarchy to generate every single possible value of $\phi$, but such a task would be computationally prohibitive. Rather, in order to increase the sample size of $\phi$, it is enough to use Monte Carlo methods to randomly select a different order of the likelihood in the Dirichlet-categorical hierarchy, toward generating Monte Carlo values of $\phi$. 

 Practical experimentation with the datasets indicated that using a Gaussian distribution was not always the output. Upon visual inspection of the histograms and kernel density estimates of their distribution, it was clear that using a Gaussian distribution to model the credible region would not necessarily bring about the most accurate results. In several instances the Monte Carlo method of approximation resulted in skewed distributions whose boundaries would not meet well assuming the standard error. Accordingly, rather than use the traditional method of Monte Carlo estimation where the mean
\begin{equation*}
E[\phi] = \frac{1}{N}\sum_{i=1}^{N} \phi_i
\end{equation*}
is the estimator of $\hat{\phi}$ \citep[51]{hammersley_monte_1979}, I decided to use the mode, where $\hat{\phi} = \arg\max_{\phi} p(\phi)$ of the densities of $\phi$. Similarly the credible region around the Monte Carlo estimation of $\hat{\phi}$ was defined as a subset $C$ of values of $\phi$ such that
\begin{equation*}
1- \alpha \leq P(C) = \int_{C} p(\phi) d\phi
\end{equation*}
where $(1-\alpha) \times 100$ represents the level of probability that $\phi$ lies within $C$ \citetext{\citealt{jaynes_confidence_1976}; \citealt[48]{carlin_bayesian_2009}}. Given the skewness and variance in densities of $\phi$, the highest posterior density interval was used to calculate the credible region, in order to obtain the most probable values of $\phi$, where $\alpha = 0.1$, resulting in a 90\% credible region.

\subsection{Additional Variables: Time and Quantity}

The example above illustrates how to sample a measure $\phi$, defined as the Hellinger distance, from probability distributions generated through a hierarchical chain of Dirichlet-categorical inference. Uncertainty about the point estimate $\hat{\phi}$ owing to non-exchangeability and sample size can be directly incorporated through Monte Carlo methods, creating random permutations of the observation of the data, resulting in a credible region.

Now it remains to bring additional considerations to bear on the model. First is the factor of time. Since the goal is to compare assemblages over time, let $j$ refer to a particular time-interval, here, a decade, such that $j_1  \equiv$ ca. 200-190 BCE, \ldots, $j_{22} \equiv$ ca. 10-20 CE. Discrete time intervals were easier to model than treating time as a continuum, and acceptable given the vagueness around dating. A Dirichlet-categorical probability distribution $p$ was designated to each time interval $j$. In other words, rather than just having one urn, we can imagine 22 urns, one for each time-period, with the estimate $\hat{\phi}_j$ between a given $p_j$ and a given $q_j$ being the value of interest for each $j$.

The dating quantities of archaeological vessels can be rendered as the expected value of a  uniform discrete distribution \citetext{\citealt[90-1]{buck_bayesian_1996}; \citealt{robertson_spatial_1999};  \citealt{fentress_counting_1987}}. For a vessel of category $i$ which has a date range from period $j_a$ to $j_b$, the value of that vessel that belongs to a time-interval $j$ is the expected value $E[x_i]_{j} = \frac{x_i}{(|j_b-j_a|+1)}$, where $x_i$ is the quantity of find (sherd count, contained in the database), and $j_a$ and $j_b$ are time-intervals provided respectively by the start and end dates in the database. Therefore, for a particular vessel category $i$ (where the total number of categories is $K$) in a time-interval $j$, $c_{i,j} =  E[x_i]_{j}$ represents the quantity of the draw for that category $i$, while $n_j = \sum_i E[x_i]_j$, the total number of all quantities drawn within the time interval $j$. Returning to the categorical likelihood above, $l(\boldsymbol{\theta ; x})$ and the calculation of the posterior probability $p(\boldsymbol{\theta | x})$, this will entail that for a given time-interval $j$,
 \begin{equation*}
 E[\theta_{i,j}|x_{i,j}] = \frac{\alpha_{i,j} + c_{i,j}}{\sum_i \alpha_{i,j} + n_j}.
 \end{equation*}
 
 To recapitulate, the frequency of vessels over time from a given project (or any context) $m$ can be summarily expressed as a set of $j$ probability distributions $p_{m,j}(\boldsymbol{\theta_j|x_j})$. Construing this process in a Bayesian framework appeals to  common sense, in the way in which estimations change on the basis of new information. The last posterior estimate is, after all, another prior, which can be updated with another likelihood. Bayesian inference moreover allows for the effects of sample size to be directly incorporated into the estimation of the value of interest, the degree of difference between the two distributions, and expression of the strength of certainty in that estimate.

	\subsection{Scaling Context: ``Combining'' Distributions}\label{sec:context}
	
In the above model, the unit of context, $m$, is an urn. Context is scalable, however, and can be construed as an archaeological layer, phase, area, or even entire site or region. In the set of data used to illustrate this method, $m$ was defined as the archaeological project itself, whether site or survey. Yet it is clear that the question of context raises an important issue, which is how to conform the above urn model to the case of archaeological finds collection where context can be variably defined. Just as it is easy to redefine sets of data into small contexts, the intuition behind combining contexts, casting the draws together from multiple urns, requires some comment.

If one merely aggregates the pulled items from two different urns, $m_1$ and $m_2$, the quantity of draws (the number of times the Dirichlet-categorical inference has occurred) will affect the resulting distribution: in other words, if more data has been observed for $m_1$ than $m_2$, the probability distribution which views $m_1$ and $m_2$ as part of the same sampling context will more closely represent $m_1$ than $m_2$. While it would be advisable to draw the amount of objects from each urn to evenly represent each sampling context, this is not always an optimal solution, as meaningful information about the sampling distributions could be needlessly lost.

Some sort of ``weighting'' mechanism would be advisable to ensure that an over-accumulation of data from one context (in comparison to under-accumulation from another) is not affecting the estimates of $\phi$ if combining from different contexts. Yet, the motivation behind this weighting mechanism ought to follow upon a motivated criterion. On the one hand, there might be situations in which aggregating the finds of one urn with another without regard to the number of draws (in an ``unweighted'' manner) are not necessarily problematic. On the other hand, there might be situations where ``weighting'' each urn's draws evenly might be problematic, and instead where certain urns (certain contexts) \textit{ought} to be more heavily weighted. In a regional aggregation of finds, one could insist upon weighting finds through \textit{per capita} estimates of other factors, like estimated populations, in order to accord greater weight to larger population centers. 

This criterion therefore should be left open, but in this case, weighting each individual project proportionally is viewed primarily as a way to mitigate the effect of project intensity on estimates. More intensive projects that have conducted fieldwork for a longer period of time will produce more data. Therefore, weighting counts proportionally to their urn populations when ``combining'' sources to construct a regional-level distribution, for $m_d$, for $d = 1, \ldots, D$ different distributions ($D$ is the total number of contexts to be combined) a weighted
 probability  distribution $p(\boldsymbol{\theta | x})$ can be obtained by the expression,
\begin{equation*}
 E[\theta_{i,j}|x_{i,j}] =  \frac{1}{D} \sum_{d=1}^D p_d({\theta_{i,j}|x_{i,j}}) = \frac{1}{D} \sum_{d=1}^D \frac{\alpha_{i,j,d} + c_{i,j,d}}{\sum_i \alpha_{i,j,d} + n_{j,d}} 
 \end{equation*}
Weighting each distribution thus comprises an additional step after each draw from an urn, but before the calculation of $\hat{\phi}$. 

\section{Implementation of the Model}

\begin{figure}[p!]
	\centering
	\includegraphics[width=1\textwidth]{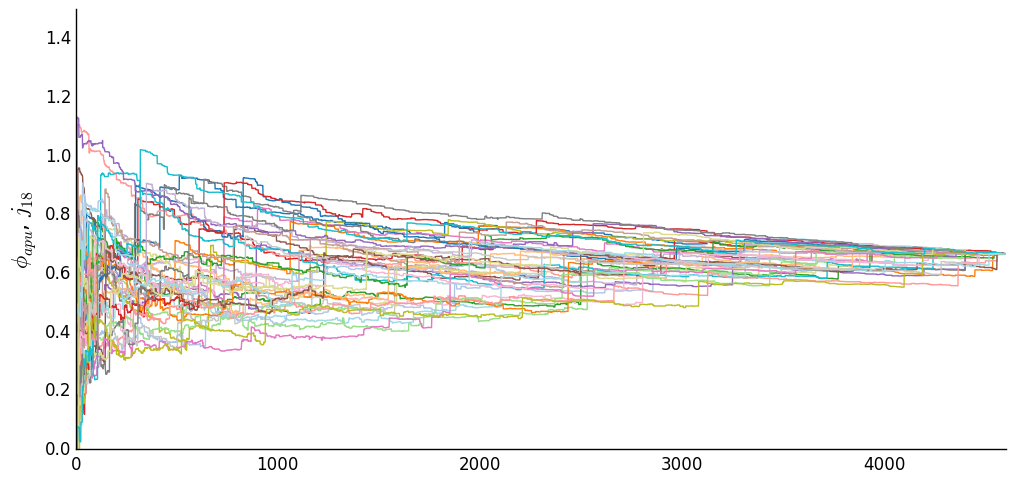}
	\caption{Example of 40 different permutations of the order of the records from the database, in order to generate Monte Carlo estimates of $\phi$, unweighted, representing the Hellinger distance between the Apulian and Italian data where $j = 18$.\label{phiconvergence1}}
\end{figure}

\begin{figure}[p!]
	\centering

			\includegraphics[width=0.7\textwidth]{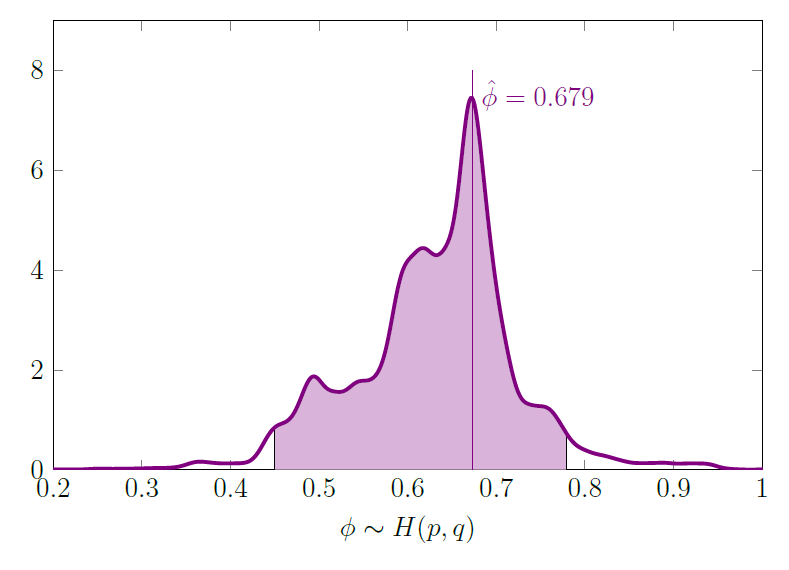}

	\caption{The kernel density estimate of the values of $\phi$ illustrated in Fig. \ref{phiconvergence1}, showing the maximum estimate $\hat{\phi}$, the mode, and the highest posterior density estimate to obtain a 90\% credible region, shaded in purple.\label{phiconvergence1a}}
\end{figure}

\begin{figure}[p!]
	\centering
	\includegraphics[width=1\textwidth]{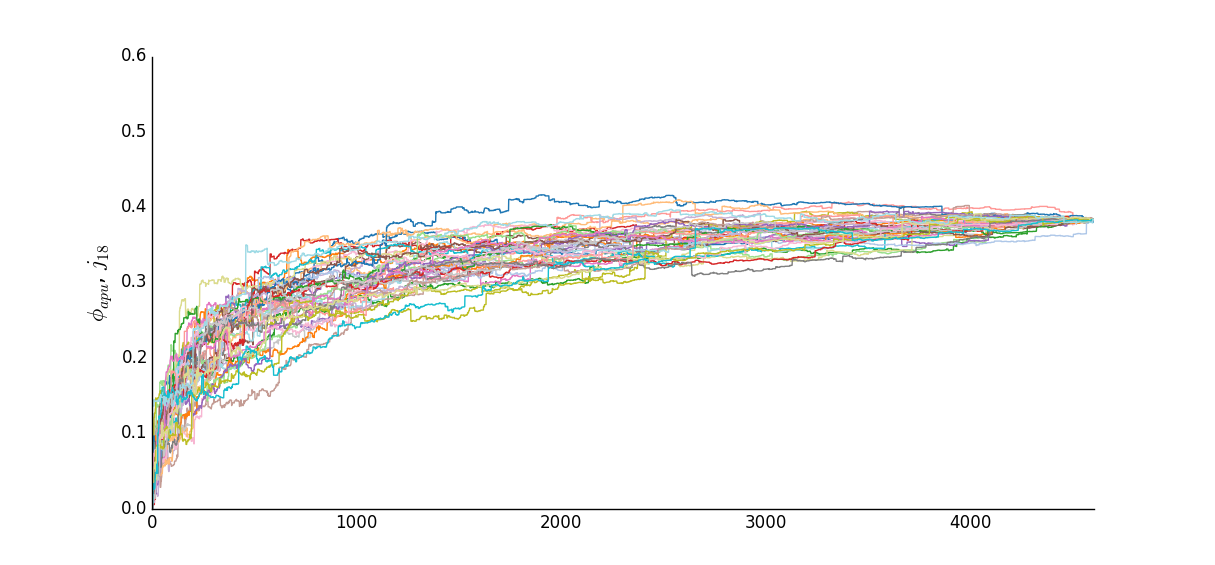}
	\caption{Example of 40 different permutations of the order of the records from the database, in order to generate Monte Carlo estimates of $\phi$, weighted, representing the Hellinger distance between the Apulian and Italian data where $j = 18$.\label{phiconvergence2}}
\end{figure}

\begin{figure}[p!]
	\centering
			\includegraphics[width=0.7\textwidth]{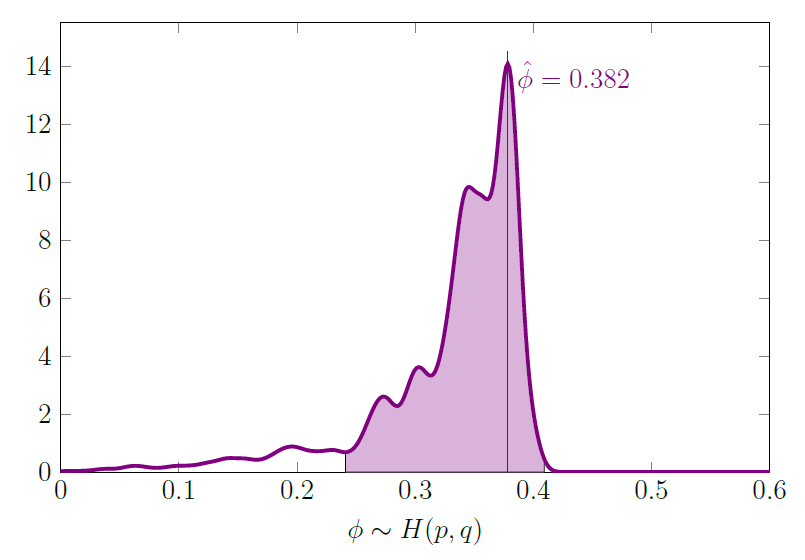}
	\caption{The kernel density estimate of the values of $\phi$ illustrated in Fig. \ref{phiconvergence2}, showing the maximum estimate $\hat{\phi}$, the mode, and the highest posterior density estimate to obtain a 90\% credible region, shaded in purple.\label{phiconvergence2a}}
\end{figure}

Implementing the above model on the set of data consisted of a python script which would pull records from the database (in effect, pulling items from an urn), using a pseudorandom number generator to carry out the Monte Carlo permutation of $\hat{\phi}$ 40 times, to select the records in different orders. Out of a total of 4,602 records (totalling a sherd count 29,466) under the parameterization by vessel form, and 4,865 (sherd count of 33,017) by vessel ware (since some vessels were unclassifiable only in either ware or form), this resulted in a Monte Carlo sample size of $N_1 = 184,080$ for $\phi$ under the paramaterization by vessel form 
and $N_2 = 194,600$ under the parameterization by vessel ware. 
Thus, a Monte Carlo approximation of $\hat{\phi}$ was obtained for each parameterization, in each time-interval, from 200 BCE to 20 CE, for each region. Figures \ref{phiconvergence1} and \ref{phiconvergence2} illustrate an overlay of 40 different orderings for the likelihood, showing the generation of Monte Carlo values of $\phi$, as well as their kernel density estimates (Figs. \ref{phiconvergence1a} and \ref{phiconvergence2a}).


Taking the categorical distribution of the different wares and forms of ceramic and glass vessels as a snapshot of the habits of food and drink consumption, the goal was then to construct three regional distributions for Etruria, Latium, and Apulia. Both unweighted and weighted methods were used to draw data from these projects, which resulted in different estimates of $\hat{\phi}$. Thus, there were three different regional probability distributions, $p_{\mathrm{etr}}$, $p_{\mathrm{lat}}$, and $p_{\mathrm{apu}}$, according to both unweighted and weighted methods.

\begin{figure}[t!]
	\centering
			\includegraphics[width=0.25\textwidth]{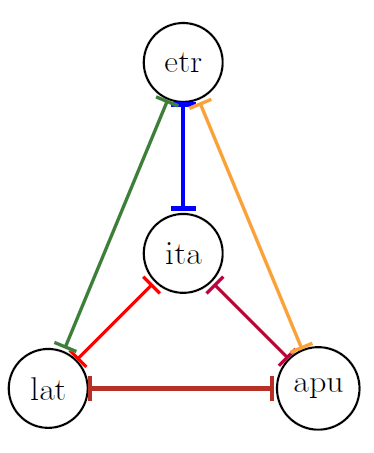}
	\caption{The construction of interregional comparisons. The Helligner distances between Etruria, Latium, and Apulia are compared against an ``Italian'' dataset (comprised of all assemblages), and also  one other.\label{fig:regdiagram}}
\end{figure}

Calculating regional difference was accomplished by realizing a global Italian distribution from the entirety of the dataset, $p_{\mathrm{ita}}$. Regionalism can be construed as deviation from a global norm, an abstracted ``pan-Italian'' habit of eating and drinking created from the entire assemblage of material in the database. Thus, each regional probability distribution was measured for its local distinctiveness or generic quality, in the degree to which it imitated or deviated from  a generic Italian distribution. This distribution is merely an abstraction to serve as the benchmark of regional variation.\footnote{Measuring $p_{\mathrm{etr}}$, $p_{\mathrm{lat}}$, and $p_{\mathrm{apu}}$ against $p_{\mathrm{ita}}$ in this way was done initially owing to the use of the Kullback-Leibler divergence, and since $D_{KL}(p||q) \neq D_{KL}(q||p)$, a probability distribution $q$ had to be used which was consistent across both measures. See note \ref{note1} on page \pageref{note1}.} In addition, the Hellinger distance also allows for the direct comparison of one assemblage with another, such that three more comparisons can be made, each between $p_{\mathrm{etr}}$, $p_{\mathrm{lat}}$, and $p_{\mathrm{apu}}$ (Fig. \ref{fig:regdiagram}).

Finally, regarding the initial value of the prior $p(\boldsymbol{\theta})$, $\boldsymbol{\alpha}$, each $\alpha_{i,j}$was taken as a flat prior, 1, such that $\sum_{i=1}^{K} \alpha_{i,j} = K$.

	\section{Results and Discussion}
	
\begin{figure}[p!]
\centering
\includegraphics[width=1\textwidth]{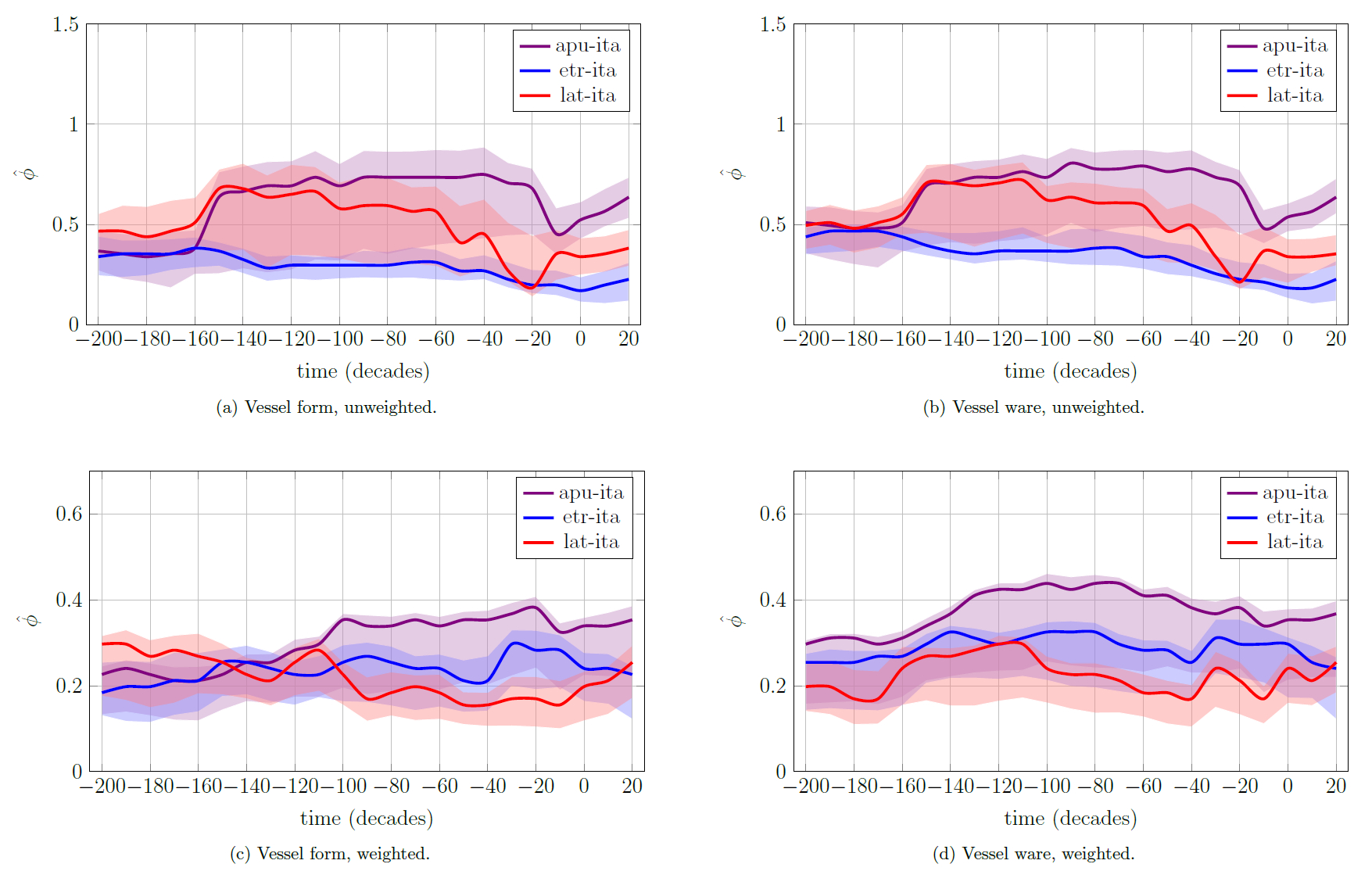}
\caption{Hellinger distance for regions from the Italian mean. Higher values indicate greater distance from the mean, and hence greater differentiation from what would be ``typical'' Italian consumer habits.\label{fig:italyresults}}
\end{figure}

\begin{figure}[p!]
\centering
\includegraphics[width=1\textwidth]{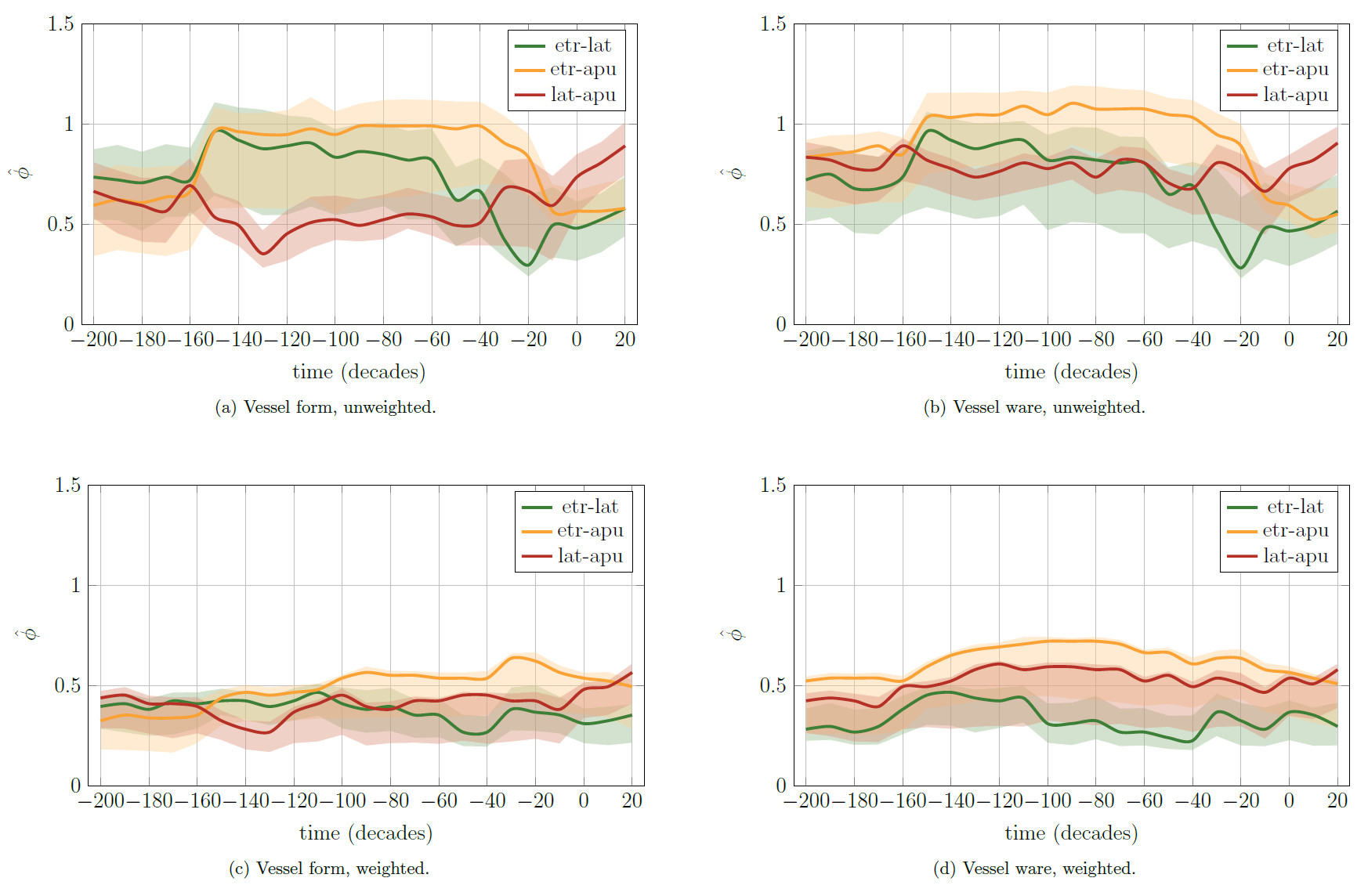}
\caption{Hellinger distances between each of the regions, with 90\% credible region. Higher values indicate greater distance between the two assemblages in terms of their total composition.\label{fig:interregresults}}
\end{figure}

The results here are grouped into two sets for the period from 200 BCE to 20 CE. The first uses a baseline comparison of a constructed Italian norm (Fig. \ref{fig:italyresults}), the second illustrates the degree of differentiation from one region to the next (Fig. \ref{fig:interregresults}). Within each of these two sets are presented the Hellinger distance calculated using different parameterizations (by vessel form and ware) and by different methods (using an unweighted and weighted set of data). Comparing these outputs illustrates the effects when using different categories on the same evidence, and of different methods of weighting. Again, using unweighted distributions can be viewed as letting intensity of fieldwork have an effect upon the resulting counts. The graphs thus represent the degree of difference between each region's habits in the consumption of food, and allow us to assess whether or not those particular lifestyle patterns  are becoming more unified or not.

In the case of measuring each region against a constructed Italian norm  (Fig. \ref{fig:italyresults}), the case of using unweighted data does not appear to alter the results between choosing vessel ware and form: the same trends are largely observable in either parameterization. On the one hand this could be due to the fact that the two parameterizations are not wholly independent of one another: vessels such as transport amphorae and lamps were unique categories in either parameterization. This result should not be taken to apply as a general rule: within certain assemblages, vessel morphology and function, taken entirely independent of manufacturing technique, could illustrate different patterns of habit from vessel ware.

	Computing regional distributions either through the raw data or through probability distributions generated from each project---in other words, using ``unweighted'' versus ``weighted'' methods---did significantly alter the results. On the one hand, giving each project an equal effect in he regional distribution raises further questions about how the data should be weighted (how to weight multiple projects within the same city, for example). Yet on the other hand the factor of intensity of fieldwork and the extent of publication, to say nothing of the quantity of material deposited in urban contexts versus rural ones, has a noticeable impact on the results. Due to the large amount of data recovered from Etrurian sites, for example, the distance of Etruria is much closer to the Italian norm throughout the study period, which reflects the view that Etrurian assemblages are closer to the norm simply because they comprise a higher quantity of the total collection to begin with. By considering regional distributions from the probability distributions of each project, however, that influence is negated, revealing that Latial customs are for the most part, though not always, closer to a generic ``Italian'' mode of food consumption.

	In terms of interpreting the results, the method of using unweighted distributions illustrates a situation in which regional variability in the norms of food consumption are apparent starting from the middle of the second century BCE, in the case of Latium and Apulia, and that those norms reconverge toward Italian norms at different rates. Given the 90\% credible regions around the Monte Carlo estimate of $\hat{\phi}$, such a trend would seem securely identifiable in the second half of the first century BCE and onward: notice should be taken of the overlap in the interval for the case of Latium and Apulia. Nor, it would seem, is such distinction permanent: a degree of increasing differentiation again emerges in the fist century CE in the case of the Apulian distribution. Given the bounded space of the Hellinger distance, which resides in $[0,\sqrt{2}]$, these trends are all occurring more closely to the Italian norm: given the synthesis of the parameterizations of vessel ware and form, this is not surprising.

	In the case where the distributions have been normalized by each project, the resulting trends present an alternative perspective. In the situation where vessel form is used as the parameterization, clear moments of shifting trends in regional differentiation are hard to extricate throughout much of the period of the late Republic. At least until the mid- to late first century BCE, the modes and credible regions overlap to a significant degree. At last in the case of the Apulian group, a level of distinction is observed at a higher rate, at least with 90\% certainty, than that of the Latial group, pointing to continued trends of cultural differentiation throughout the first century BCE from the Italian norm. The situation in which vessel ware is used as the parameterization illustrates a perdurance of regional distinction throughout the last two centuries BCE at different levels.
	
	Yet, one could argue that the framing of regional differentiation through comparison with the global mean cannot be substantiated on the basis of a constructed Italian set of habits, since it could be problematic to assume that all Italy serves as a basis for comparison. In this light, the symmetry of the Hellinger distance proves useful, as it allows each region to be compared with another, without reference to an ``expected'' value, which would be necessary in the case of the chi-squared distance.

	As above, unweighted counts produced more striking patterns of regional variation, largely similar notwithstanding whether vessel ware or form is used as the parameterization (Fig. \ref{fig:interregresults}). While Etruria becomes more similar to Latium and to Apulia in its habits, the difference between Latium and Apulia in fact appears to increase starting in the later part of the first century BCE. Weighted counts on the other hand produced patterns that were far less drastic, with the difference between Latium and Apulia visible in the case of vessel ware, but not in vessel morphology: the differences between Etruria and the other two regions on the whole appear rather stable over time. On the whole it is worthwhile to note that Etruria and Latium bear the most similarity over the last two centuries BCE in terms of vessel morphology, with Latium and Apulia the next most similar, and Etruria and Apulia the least similar.
	
	Weighted counts produce trends which are much more stable throughout the period 200 BCE to 20 CE, with the different between Latium and Apulia increasing first in the mid-second century BCE, and again in the last decade of the first century BCE, when examining the use of vessels in terms of their ware. In terms of vessel form, the level of regional differentiation is fairly uniform over the last two centuries BCE as well, yet with the mid-second century BCE emerging again as a period in which all inter-regional differences increase. No one region is deviating from the others to any greater or lesser extent according to vessel ware, but rather all become more slightly more distinct, starting in the decade of the 160s BCE. This assessment corroborates the pattern when comparing against the global Italian mean, that a period of differentiation of vessel class and form emerged among the regions of Etruria, Latium, and Apulia starting in the mid-second century BCE. The Augustan period of the late first century BCE shows that while Etruria and Apulia became slightly more similar in their habits, Latium and Apulia nevertheless became more distinct, both from one another, and from the constructed Italian norm.
	

	\section{Conclusions}
	
	The purpose of this paper has been to demonstrate the use of a Bayesian framework of inference to establish quantitative comparisons of archaeological assemblages, as well as to establish a credible interval around those estimates. The data used to illustrate this method thus provide a first step toward broader comparisons, and it should not be thought that such results provide a definitive statement on regional differentiation in consumer practices of food in Republican Italy. Not just additional incorporation of vessel assemblages from a greater array of contexts, but even correlating those comparisons with broadly collected data from both paleoethnobotany \citep{mercuri_pollen_2015} and zooarchaeology \citep{mackinnon_production_2004}, are essential for establishing more a more holistic picture of the practices of ancient food consumption, in a way that allows for measurable differentiation from society to society.
	
	Nevertheless, the above method allows for the synthesis and comparison of  data drawn from a variety of sources, toward comparing material assemblages in their capacity as a mechanism of habitual action. In this case, focusing on the way in which habits centered around food speak to broader social change, the categorical distribution of different vessels that contributed to the shipment, processing, and consumption of food have been taken as an intrinsic component of the way which the inhabitants of ancient Italy ate and drank. Whereas the contents of those meals and full reconstruction of those habits are inconceivable even to imagine---the thousands of meals over the course of centuries---discernment of how those communities' habits varied is an essential step in our own reconstruction of the significance behind behavior in antiquity. Whether different modes of habitual action experienced the regular flow of long-term changes as constructs of fashion and generational transience, or whether there were abrupt and sudden departures in the cultural code, such developments can be quantified and measured.

In this way, theoretical discourse regarding the question of cultural change in Republican and Augustan Italy can avail itself of a deeper use of material evidence, establishing comparisons from the chaos of material waste, as here, with vessels, in the case of the habits and regimens of food consumption for studying mass society. Consideration of social and economic change need not proceed from \textit{a priori}  assumptions about the dynamics of cultural change in an imperial context, but rather from the evidence itself, building from the data up. Already, the pilot dataset assembled in this study suggest that a steady trend of cultural differentiation in these habits from the mid-second century BCE onward, and which the period of Augustus did little to affect or alter, or did so only temporarily.

The results of this pilot study moreover indicate that regional distinctions in the habits of eating and drinking  proceed according to their own logic and are not necessarily tied to singular causes or events, and that alterations in the fashions of those behaviors occur due to complex reasons which are ordered and are subject to change. Accordingly, they warrant investigation, both in their causes and effects. Like the construction and use of language, the history of the formation and development of consumer habits constitute a key component of past experience, and, in the context of material remains, such practices are the motor that generates the archaeological record itself.

\section{Acknowledgments}
	
Aspects of this paper were presented at the Theoretical Roman Archaeology Conference Workshop, ``Making practice perfect: approaches to everyday life in Roman archaeology,'' held at the University College London Institute of Archaeology on 30 January 2016.  This paper represents an extension of the method developed in my Ph.D. dissertation, and I would therefore like to take this opportunity to express my gratitude to my committee, Nancy de Grummond, David Stone, Daniel Pullen, John Marincola, Debajyoti Sinha, and David Levenson for their kind support in that process, as well as Kim Bowes for her generous encouragement of this work. The  dissertation was also supported by a Fulbright Full Grant to conduct research at the Universit\`a di Siena from 2012 to 2013, and it is my pleasure to thank Franco Cambi for his warm welcome and support, Emanuele Vaccaro and my Italian friends and colleagues for their help, as well as the US-Italy Fulbright Commission. Development of the probabilistic model and coding would not have been possible without the generous award of a fellowship at the Tennessee Humanities Center at the University of Tennessee, Knoxville, from 2015 to 2016. I would also like to extend my deepest thanks to the anonymous reviewers, whose comments have greatly improved the quality of this paper. Finally, my wife, Jenny, has my eternal gratitude for her patience with me throughout our odyssey together. All errors remain my own.

\appendix

\section{Database Overview} \label{appendix1}
	
Data were entered by the author into an SQLite database which was queried in \texttt{python}, also available at {https://github.com/scollinselliott/hellinger-montecarlo}. Owing to the fact that presentation and content of these archaeological data were prepared in a heterogeneous manner, they had to be standardized.

Each tuple in the  database is assigned a unique nine-digit \texttt{key}, starting with the three-digit project code. This sometimes involved breaking apart records in the published literature, as they would include counts of rims, bases, and body sherds within their same record.   The citation field, \texttt{project}, contains a three-digit code (Table \ref{bibl1}) which corresponds to the site or survey which produced the particular find. The field \texttt{bibl2} records the page, figure, or, in the case of electronic formats of ceramic record sheets, the line number of the entry.
	
	\begin{table}[t!]
		\centering
		\begin{tabular}{ll}
			\toprule
			Field		&	Description\\
			\midrule
			\texttt{key}			&	Unique Key\\
			\texttt{project}		&	Context / citation for record\\
			\texttt{bibl2}		&	Page and/or figure of record\\
			\texttt{frtype}		&	Type of sherd (rim, base, etc.)\\
			\texttt{nfr}			&	Quantity (sherd count)\\		
			\texttt{nw}			&	Quantity (weight)\\
			\texttt{sem1}		&	Semantic set for vessel form (morphological-functional)\\
			\texttt{sem2}		&	Semantic set for vessel ware (technical-functional)\\
			\texttt{kat1}		&	Vessel form (synthkat determined)\\
			\texttt{kat2}		&	Vessel ware (synthkat determined)\\
			\texttt{date1}		&	Start date (Likely \textit{terminus post quem})\\
			\texttt{date2}		&	End date (Likely \textit{terminus ante quem})\\
			\bottomrule		
		\end{tabular}
		\caption{Database fields for the pilot dataset.\label{databasefields}}
	\end{table}
	
The field \texttt{frtype}, which was not used in this study, records the type of fragment: 0 = nonid, 1 = rim, 2 = handle, 3 = base, 4 = body sherd, 5 = shoulder (lamp only), 6 = spout (lamp only). Quantification is presented in the fields of \texttt{nfr}, sherd count, and \texttt{nw}, sherd weight. Originally, an EVE field was included, but owing to the dearth of EVEs in Roman archaeology in Italy, it was dropped. The aim behind stipulating vessel fragment types and different quantification schemes was to allow for greater options in choosing which data are subject to analysis, owing to the fact that it is ideal for multiple methods of quantification to be used: sherd count, weight, and EVE being the most useful \citep{orton_quantitative_1975,orton_how_1993,orton_`four_2009,orton_statistical_1990}. Vessel weights were however not always provided. Hence, sherd count alone was used in this study. In the case of instances which reported a single vessel, that instance was taken as representing one sherd, a stop-gap measure to ensure its inclusion in the database. Using different quantification schemes simultaneously would ameliorate biases in the other methods. When precise quantitative data was lacking but presence was documented, or specified only in qualitative terms (such as ``many'' or ``several''), an arbitrary base quantity of 1 was entered into \texttt{nfr} as a placeholder, to at least count for a minimum value. 
	
The fields \texttt{date1} and \texttt{date2} contained the record's \textit{terminus post quem} and \textit{ante quem}, respectively, which were used to provide a uniform distribution over time to date the vessels. These dates were those assigned in the published record, not always predicted upon the ware itself but rather the dating of the artifact in its particular context.

\bibliographystyle{elsarticle-harv} 
\bibliography{refs}

\end{document}